\documentclass[nofootinbib,unsortedaddress]{revtex4}
\usepackage{amsmath}
\usepackage{dcolumn}
\usepackage{bm}
\usepackage{xcolor}
\usepackage{url}
\usepackage{graphicx}
\usepackage{float}

\usepackage[letterpaper,left=2.0cm,right=2.0cm,top=2.5cm,bottom=2.5cm]{geometry} 
\usepackage{siunitx} 
\usepackage{import} 
\usepackage{amssymb}
\usepackage{subcaption}

\usepackage[english]{babel}
\usepackage{hyperref}
\hypersetup{
    colorlinks=true,
    linkcolor=blue,
    filecolor=magenta,      
    urlcolor=blue,
    citecolor=blue,
}

\begin{document}

\title{Null Raychaudhuri Equation and the Impossibility of Traversable Wormholes in Unimodular Gravity}

\author{Erick Past\'en}
\email{erick.contreras@usm.cl}
\thanks{Corresponding author}
\affiliation{Departamento de F\'{\i}sica, Universidad de Santiago de Chile,
Avenida V\'{\i}ctor Jara 3493, Estaci\'on Central, 9170124, Santiago, Chile}

\author{Marco Bosquez}
\email{marco.bosquez@usach.cl}
\affiliation{Instituto de F\'{\i}sica, Pontificia Universidad Cat{\'o}lica de Valpara\'{\i}so,
Avenida Universidad 330, Curauma, Valparaíso, Chile.}

\author{Norman Cruz}
\email{norman.cruz@usach.cl}
\affiliation{Departamento de F\'{\i}sica, Universidad de Santiago de Chile,
Avenida V\'{\i}ctor Jara 3493, Estaci\'on Central, 9170124, Santiago, Chile}


\begin{abstract}
We formulate wormhole traversability as a purely local and covariant null defocusing condition derived from the Raychaudhuri equation, independently of any specific metric ansatz. Since unimodular gravity preserves the local affine and causal structure of spacetime, the null focusing properties of geodesic congruences remain identical to those of general relativity. We prove that any genuinely traversable wormhole in unimodular gravity necessarily violates the null energy condition, establishing a general local no--go theorem for wormholes supported by ordinary matter in this framework.
\end{abstract}


\maketitle

\section{Introduction}

The Raychaudhuri equation occupies a central role in general relativity as a purely geometric identity
governing the focusing properties of families of geodesics \citep{Raychaudhuri1955}.
Together with appropriate causality and energy assumptions, it underlies some of the deepest results
of the theory, including the Hawking--Penrose singularity theorems
\citep{Penrose1965,Hawking1966}, and provides a covariant description of gravitational attraction in
terms of geodesic convergence \citep{mtw1973gravitation,wald1984general,ellis2012relativistic}.
A remarkable feature of the Raychaudhuri equation is that it is entirely independent of the gravitational
field equations: it relies only on the affine structure of spacetime and the definition of curvature
through the Riemann tensor. 
However, physical statements about attractive gravity, geodesic focusing, or energy conditions arise only after relating the Ricci tensor to matter through a specific set of gravitational field equations. In this sense, the Raychaudhuri equation by itself only describes the kinematical evolution of geodesic congruences, whereas the physical origin of focusing depends on the dynamical sector of the gravitational theory under consideration. Consequently, although the null Raychaudhuri equation remains purely geometric, it is not a priori obvious that different gravitational theories must lead to identical conclusions regarding geodesic focusing or constraints on the admissible matter content of spacetime.

Traversable wormholes provide a natural arena in which the interplay between geometry, causality and
energy conditions becomes particularly transparent.
Morris and Thorne \citep{MorrisThorne1988} showed that static, spherically symmetric
wormholes can be constructed at the price of violating the classical energy conditions, a conclusion
further sharpened in \citep{MorrisThorneYurtsever1988}.
Subsequent analyses clarified that the essential obstruction to traversability is not tied to a
specific metric ansatz, but rather to the focusing of null geodesic congruences threading the wormhole
throat.
In particular, \citep{HochbergVisser1997} reformulated the flare--out condition in
geometric terms, relating traversability to the behaviour of null generators and the absence of
caustic formation.

Unimodular gravity is a modification of general relativity obtained by restricting the variational
principle to metrics with fixed determinant \citep{Unruh1989,Alvarez2006,Padilla2015}.
Although dynamically equivalent to general relativity up to an integration constant that plays the
role of a cosmological constant, unimodular gravity has attracted renewed attention in recent years,
particularly in discussions of vacuum energy and the cosmological constant. 
Because unimodular gravity modifies the trace sector of Einstein's equations, the theory allows vacuum-energy-like contributions to emerge as integration constants rather than as direct sources in the action. This feature has motivated extensive discussions regarding the cosmological constant problem and the interpretation of effective stress--energy sources in curved spacetimes \citep{Ellis2011,Padilla2015,Alvarez2006}. In the context of wormhole physics, these differences raised the possibility that the usual connection between traversability and exotic matter could be altered, potentially allowing traversable configurations supported by ordinary matter satisfying the null energy condition.
Motivated by these differences at the level of the field equations, it has recently been claimed that
unimodular gravity may admit traversable wormhole solutions supported by ordinary matter satisfying
the null energy condition \citep{Agrawal_2023}.
Related critiques have argued that such conclusions are, at least, questionable in the particular case of the Morris--Thorne metric \citep{Cataldo2025}.

The purpose of this work is to revisit the issue of traversable wormholes in unimodular gravity from a
purely geometric and local perspective.
Rather than focusing on particular solutions or effective energy--momentum tensors, we formulate the
condition for traversability directly in terms of the null Raychaudhuri equation and the behaviour of
null geodesic congruences at the throat.
We show that unimodular gravity shares the same local affine and causal structure as general relativity,
so that the null Raychaudhuri equation remains unchanged.
As a consequence, the requirement of null defocusing at a traversable throat implies violation of the
null energy condition independently of whether the underlying field equations are those of general
relativity or unimodular gravity.

This analysis establishes a local no--go theorem for genuinely traversable wormholes in unimodular
gravity supported by ordinary matter, and clarifies the precise geometric sense in which such claims must be interpreted.

\section{Null Raychaudhuri equation}
\label{sec:NRE}

Raychaudhuri equation \citep{Raychaudhuri1955} is a field--equation independent local identity that follows from the definition of a congruence, the Ricci identity, and the kinematical decomposition of the deformation tensor.
It has been used extensively to reveal deep properties of general relativity, such as the Hawking--Penrose singularity theorems \citep{Hawking1966,Penrose1965}.

Let $u^a$ be the unit tangent field to a timelike congruence, $u^a u_a=-1$. The spatial projector orthogonal
to $u^a$ is
\begin{equation}
h_{ab} = g_{ab} + u_a u_b,
\qquad
h_{ab}u^b=0.
\end{equation}
The \textit{deformation tensor} can be defined as the 4-gradient of the unit tangent field as
\begin{equation}
B_{ab} = \nabla_b u_a.
\end{equation}
Its standard kinematical decomposition reads
\begin{equation}
\nabla_b u_a
= \frac{1}{3}\,\theta\, h_{ab} + \sigma_{ab} + \omega_{ab} - u_a a_b,
\label{eq:kin_decomp_timelike}
\end{equation}
where $\theta=\nabla_a u^a$ is the expansion scalar, $a^a=u^b\nabla_b u^a$ is the 4-acceleration,
$\sigma_{ab}$ (shear) is the symmetric, trace-free and spatial part of the tensor while $\omega_{ab}$ (vorticity) is antisymmetric and spatial.

The evolution of the expansion along the flow can be computed using the Ricci identity,
\begin{equation}
(\nabla_c\nabla_b-\nabla_b\nabla_c)u_a = R_{adcb}u^d,
\label{eq:ricci_identity}
\end{equation}
contracting with $u^c$ and using the multiplication rules for derivatives one finds
\begin{align}
u^c\nabla_c(\nabla_b u_a)
&= \nabla_b(u^c\nabla_c u_a) - (\nabla_b u^c)(\nabla_c u_a) + R_{adcb}u^d u^c \nonumber\\
&= \nabla_b a_a - B^c{}_b B_{ac} + R_{adcb}u^d u^c.
\end{align}
Taking the trace with $g^{ab}$ gives
\begin{equation}
\dot\theta \equiv u^a\nabla_a\theta
= \nabla_a a^a - B_{ab}B^{ba} + g^{ab}R_{adcb}u^d u^c.
\end{equation}
Using the antisymmetry of the Riemann tensor in its last two indices,
\begin{equation}
R_{adcb}=-R_{adbc},
\end{equation}
together with the definition of the Ricci tensor,
\begin{equation}
R_{dc}=g^{ab}R_{adbc},
\end{equation}
one obtains
\begin{equation}
g^{ab}R_{adcb}u^d u^c
=
- R_{dc}u^d u^c
=
- R_{ab}u^a u^b.
\end{equation}
Therefore,
\begin{equation}
\dot\theta \equiv u^a\nabla_a\theta
= \nabla_a a^a - B_{ab}B^{ba} - R_{ab}u^a u^b.
\label{eq:theta_trace_intermediate}
\end{equation}
Finally, substituting the decomposition \eqref{eq:kin_decomp_timelike} and using standard algebra yields
\begin{equation}
B_{ab}B^{ba}
= \frac{1}{3}\theta^2 + \sigma_{ab}\sigma^{ab} - \omega_{ab}\omega^{ab},
\end{equation}
so that the Raychaudhuri equation for a timelike congruence becomes
\begin{equation}
\boxed{
\dot\theta
= -\frac{1}{3}\theta^2
- \sigma_{ab}\sigma^{ab}
+ \omega_{ab}\omega^{ab}
- R_{ab}u^a u^b
+ \nabla_a a^a
}\,.
\label{eq:raychaudhuri_timelike}
\end{equation}
For geodesic flow ($a^a=0$) this reduces to the familiar focusing form where shear and expansion always
decrease $\theta$, while vorticity counteracts focusing. For a null congruence, let $k^a$ be the tangent vector, $k^a k_a=0$. If the vector is affinely parametrised, then
$k^b\nabla_b k^a=0$. One defines the expansion $\theta$, shear $\sigma_{ab}$ and vorticity $\omega_{ab}$ by
projecting $\nabla_b k_a$ onto the two-dimensional screen space orthogonal to $k^a$ (and an auxiliary null
vector). The derivation proceeds exactly as above, yielding
\begin{equation}
\boxed{
\frac{d\theta}{d\lambda}
= -\frac{1}{2}\theta^2
- \sigma_{ab}\sigma^{ab}
+ \omega_{ab}\omega^{ab}
- R_{ab}k^a k^b
}\,.
\label{eq:raychaudhuri_null}
\end{equation}
When $\omega_{ab}=0$ and the shear term is non-negative, defocusing ($d\theta/d\lambda>0$) requires
$R_{ab}k^a k^b<0$, which is the key condition we will connect to energy conditions once a set of field equations is specified. The strong point of equation \eqref{eq:raychaudhuri_null} is that it is a purely geometric result and does not rely on any specific gravitational field equations: it only assumes the existence of a Levi--Civita connection and the standard definition of curvature through the Riemann tensor.
Unimodular gravity shares the same geometric structure as general relativity, differing only in the variational restriction imposed on the metric determinant and in the trace sector of the field equations \citep{Unruh1989,Alvarez2006,Padilla2015}.
As a consequence, the Raychaudhuri equations \eqref{eq:raychaudhuri_timelike} and
\eqref{eq:raychaudhuri_null} remain unchanged in unimodular gravity. 
Any distinction between general relativity and unimodular gravity arises only when relating the curvature term $R_{ab}k^a k^b$ to the matter content via the corresponding field equations, a point we address in the following sections.

\section{Traversability condition and flare--out}
\label{sec:flareout}

A traversable wormhole must allow causal (in particular, null) signals to pass through its throat without
encountering a caustic \footnote{
A caustic corresponds to the formation of conjugate points along the null congruence,
where neighbouring geodesics intersect and the expansion diverges to $-\infty$.
The presence of a caustic obstructs causal propagation and therefore rules out
traversability. Examples of caustic formation in general relativity include the convergence of null geodesics in gravitational collapse and in past-directed cosmological congruences approaching the initial singularity.}
. Geometrically, this requirement can be stated in terms of the behaviour of the
cross-sectional area of a congruence of null geodesics threading the throat. 
Consider an affinely parametrised null geodesic congruence with tangent $k^a$ threading a candidate throat. Let $A(\lambda)$ denote the area element of a two-dimensional screen-space
cross-section transported along the congruence, i.e.\ the transverse area orthogonal to the null generators. As the null generators evolve along the affine parameter $\lambda$, the cross-sectional area may expand or contract. The expansion scalar $\theta$ measures precisely this fractional rate of change of the transverse area along the congruence,
\begin{equation}
\theta \equiv \frac{1}{A}\frac{dA}{d\lambda}.
\label{eq:theta_area}
\end{equation}
A throat is characterised as a minimal-area cross-section along the congruence, corresponding to the point where the expansion of the null generators changes from convergence (negative expansion scalar) to divergence (positive expansion scalar). Geometrically, this implies that the transverse area becomes locally stationary at the throat, so that $\theta=0$. This requirement ensures that neighbouring null generators begin to separate after crossing the throat. Otherwise, if the cross-sectional area continued decreasing beyond the throat, the congruence would keep converging and eventually form conjugate points (caustics). If $\lambda=\lambda_0$
labels the throat, then the required condition is
\begin{equation}
\left.\frac{dA}{d\lambda}\right|_{\lambda_0}=0
\qquad\Longleftrightarrow\qquad
\theta(\lambda_0)=0.
\label{eq:throat_theta0}
\end{equation}
Traversability requires that the congruence does not form a caustic at (or immediately after) the
throat. A stationary point of the area alone is therefore not sufficient: the throat must correspond to a local minimum of $A(\lambda)$. Therefore,
\begin{equation}
\left.\frac{d^2A}{d\lambda^2}\right|_{\lambda_0} > 0.
\label{eq:throat_area_min}
\end{equation}
Differentiating Eq.~\eqref{eq:theta_area} with respect to $\lambda$, one obtains
\begin{equation}
\frac{1}{A}\frac{d^2A}{d\lambda^2} = \frac{d\theta}{d\lambda} + \theta^2,
\end{equation}
and therefore, at the throat where $\theta(\lambda_0)=0$,
\begin{equation}
\left.\frac{d^2A}{d\lambda^2}\right|_{\lambda_0} > 0
\qquad\Longleftrightarrow\qquad
\left.\frac{d\theta}{d\lambda}\right|_{\lambda_0} > 0.
\label{eq:defocusing_condition}
\end{equation}
Equation \eqref{eq:defocusing_condition} is the \emph{general flare--out/defocusing} requirement: the expansion must
increase through the minimal cross-section so that neighbouring null generators separate rather than focus. The traversability criterion adopted in this work is formulated in the language of null geodesic
congruences and the Raychaudhuri equation, and should be regarded as a local and causal condition.
It captures the minimal requirement for physical traversability, namely the absence of null focusing
and caustic formation at the throat.
While this criterion is closely related to the geometric flare--out conditions discussed in \citep{HochbergVisser1997}, it does not rely on a global variational characterization
of the throat as a minimal two--surface.

Substituting the null Raychaudhuri equation \eqref{eq:raychaudhuri_null} and evaluating at the throat
($\theta=0,\>\lambda=\lambda_0$) gives
\begin{equation}
\left.\frac{d\theta}{d\lambda}\right|_{\lambda_0}
=
-\left.\sigma_{ab}\sigma^{ab}\right|_{\lambda_0}
+\left.\omega_{ab}\omega^{ab}\right|_{\lambda_0}
-\left.R_{ab}k^a k^b\right|_{\lambda_0}.
\label{eq:raychaudhuri_at_throat}
\end{equation}
For hypersurface-orthogonal null generators (the relevant case for a static and spherically symmetric
geometry), $\omega_{ab}=0$. Since the shear term
is non-negative, $\sigma_{ab}\sigma^{ab}\ge 0$, the defocusing condition \eqref{eq:defocusing_condition} implies
the necessary inequality
\begin{equation}
\boxed{
\left.R_{ab}k^a k^b\right|_{\lambda_0} < 0
}
\qquad
(\omega_{ab}=0),
\label{eq:Ricci_defocus}
\end{equation}

In general relativity, the curvature term can be related to the stress--energy tensor through Einstein's
equations,
\begin{equation}
R_{ab}-\frac{1}{2}R\,g_{ab}+\Lambda g_{ab}=8\pi T_{ab}.
\label{eq:Einstein_eq}
\end{equation}
Contracting Eq.~\eqref{eq:Einstein_eq} with $k^a k^b$ gives
\begin{equation}
R_{ab}k^a k^b
-
\frac{1}{2}R\,g_{ab}k^a k^b
+
\Lambda g_{ab}k^a k^b
=
8\pi\,T_{ab}k^a k^b.
\end{equation}
Since the null generators satisfy
\begin{equation}
g_{ab}k^a k^b =0,
\end{equation}
both the trace and cosmological constant terms vanish, yielding
\begin{equation}
R_{ab}k^a k^b = 8\pi\,T_{ab}k^a k^b,
\label{eq:Ricci_to_Tkk}
\end{equation}
Therefore, condition \eqref{eq:Ricci_defocus} is equivalent to
\begin{equation}
\boxed{
\left.T_{ab}k^a k^b\right|_{\lambda_0} < 0,
}
\label{eq:NEC_violation}
\end{equation}

The above condition is the well-known violation of the Null Energy Condition (NEC) for traversable wormholes \citep{MorrisThorneYurtsever1988, HochbergVisser1997}. Physically, this states that any real traversable wormhole with no vorticity, requires the existence of \emph{exotic} matter.

\subsection{Morris--Thorne flare--out condition}
\label{subsec:MT_from_Raychaudhuri}

The general traversability condition derived above is purely covariant and formulated in terms of the
behaviour of null geodesic congruences. In particular, a traversable throat corresponds to a minimal-area
surface threaded by null generators satisfying
\begin{equation}
\theta=0,
\qquad
\left.\frac{d\theta}{d\lambda}\right|_{\rm throat}>0,
\label{eq:general_flareout}
\end{equation}
which guarantees defocusing and the absence of caustics. Consider now the Morris--Thorne line element \citep{MorrisThorne1988}
\begin{equation}
ds^2 = -e^{2\Phi(r)}dt^2 + \frac{dr^2}{1-b(r)/r} + r^2 d\Omega^2,
\label{eq:MT_metric}
\end{equation}
where $\Phi(r)$ is the redshift function and $b(r)$ is the shape function.
The areal radius $r$ directly determines the transverse area of spherical cross-sections,\footnote{
The interpretation of the expansion as the fractional rate of change of a transverse area,
$\theta = A^{-1} dA/d\lambda$, assumes a vorticity--free null congruence,
$\omega_{ab}=0$, so that well--defined transverse cross--sections exist.
For congruences with non--vanishing vorticity the expansion remains a local scalar quantity,
but it cannot be globally interpreted as the rate of change of the area of an integrable
two--surface.
}

\begin{equation}
A(r)=4\pi r^2.
\end{equation}
A throat is therefore identified as a stationary point of the transverse area along curves threading the geometry. Since in the Morris--Thorne parametrization the area is $A=4\pi r^2$, this condition is equivalently expressed as a minimum of the areal radius $r$ with respect to a geometrical parameter along the throat direction, such as the proper radial distance, an embedding parameter or the affine parameter. 
For a congruence parametrized by a geometrical parameter $\lambda$, the expansion is
\begin{equation}
\theta
=
\frac{1}{A}\frac{dA}{d\lambda}
=
\frac{2}{r}\frac{dr}{d\lambda}.
\end{equation}
Therefore, the throat condition $\theta=0$ implies
\begin{equation}
\left.\frac{dr}{d\lambda}\right|_{r_0}=0,
\end{equation}
i.e.\ the areal radius becomes stationary along the congruence. This condition can be made explicit by considering the embedding of an equatorial spatial slice into Euclidean space
\begin{equation}
dl^2 = dz^2 + dr^2 + r^2 d\phi^2,
\end{equation}
or
\begin{equation}
dl^2 =
\left[
1+\left(\frac{dz}{dr}\right)^2
\right]dr^2
+
r^2 d\phi^2.
\end{equation}
Therefore, the embedding profile $z(r)$ satisfies
\begin{equation}
1+\left(\frac{dz}{dr}\right)^2
=
\frac{1}{1-b(r)/r}.
\end{equation}
Equivalently,
\begin{equation}
\frac{dr}{dz}
=
\pm\left(\frac{r}{b(r)}-1\right)^{1/2}.
\end{equation}
Since the throat corresponds to a stationary point of the areal radius along the embedding geometry, one must require
\begin{equation}
\left.\frac{dr}{dz}\right|_{r_0}=0.
\end{equation}
From the previous equation this occurs when
\begin{equation}
\frac{r_0}{b(r_0)}-1=0,
\end{equation}
which implies
\begin{equation}
b(r_0)=r_0,
\label{eq:throat_MT}
\end{equation}
corresponding to $\theta=0$ in the covariant description. Evaluating the second derivative of the embedding at the throat yields
\begin{equation}
\left.\frac{d^2r}{dz^2}\right|_{r_0}
=
\left.\frac{b(r)-b'(r)\,r}{2b(r)^2}\right|_{r_0}.
\end{equation}
Requiring a flaring-out geometry, i.e. a local minimum of the areal radius with respect to the embedding parameter, implies
\begin{equation}
\left.\frac{d^2r}{dz^2}\right|_{r_0} > 0
\qquad\Longleftrightarrow\qquad r_0-b'(r_0)r_0 > 0\qquad\Longleftrightarrow\qquad
b'(r_0)<1,
\label{eq:MT_flareout}
\end{equation}
which is the familiar Morris--Thorne flare--out condition. The key point is that the inequality \eqref{eq:MT_flareout} is not an independent geometric assumption,
but rather a coordinate-specific expression of the general defocusing condition
\eqref{eq:general_flareout}. 

\section{Traversability condition in unimodular gravity}
\label{sec:UG_traversability}

In recent years, it has been claimed the existence of traversable wormhole solutions in unimodular
gravity without invoking exotic matter, i.e.\ while preserving the null energy condition (NEC) for the
matter sector \citep{Agrawal_2023}.
These results have been interpreted as evidence that unimodular gravity may evade the standard arguments that forbid traversable wormholes supported by ordinary matter in general relativity.
However, recent critiques have emphasized that the alleged avoidance of exotic matter may fail within the Morris--Thorne framework \citep{Cataldo2025}. In this section we show that the existence of traversable wormholes supported by ordinary matter in UG cannot be sustained at the level of general local null geometry. Unimodular gravity is obtained by restricting the variational principle to metrics with fixed determinant,
$\sqrt{-g}=\mathrm{const}$, leading to the traceless field equations
\begin{equation}
R_{ab} - \frac{1}{4}R\,g_{ab}
=
8\pi\left(T_{ab} - \frac{1}{4}T\,g_{ab}\right).
\label{eq:UG_field_eq}
\end{equation}
As discussed in the literature \citep{Unruh1989,Alvarez2006,Padilla2015}, unimodular gravity
differs from general relativity only in the trace sector of the field equations, while the affine structure, geodesics and curvature tensors are identical. This implies that the null Raychaudhuri equation,
\begin{equation}
\frac{d\theta}{d\lambda}
= -\frac{1}{2}\theta^2
- \sigma_{ab}\sigma^{ab}
+ \omega_{ab}\omega^{ab}
- R_{ab}k^a k^b,
\label{eq:UG_Raychaudhuri}
\end{equation}
is a purely geometric identity and therefore holds unchanged in unimodular gravity.
The traversability requirement derived in Sec.~\ref{sec:flareout} implies, for twist-free null generators,
\begin{equation}
\left.R_{ab}k^a k^b\right|_{\rm throat} < 0.
\label{eq:UG_defocus}
\end{equation}

Using the unimodular field equations \eqref{eq:UG_field_eq} and contracting with $k^a k^b$, the trace terms
again drop out due to the null character of $k^a$, yielding
\begin{equation}
R_{ab}k^a k^b = 8\pi\,T_{ab}k^a k^b,
\label{eq:UG_Rkk_Tkk}
\end{equation}
which is \emph{identical} to the corresponding relation in general relativity. Therefore, condition \eqref{eq:UG_defocus} is equivalent to
\begin{equation}
\left.T_{ab}k^a k^b\right|_{\rm throat} < 0,
\label{eq:UG_NEC_violation}
\end{equation}
i.e.\ a violation of the null energy condition along the null generators threading the throat. Any traversable wormhole supported by twist-free null congruences necessarily requires NEC violation, exactly as in general relativity.
For example, in charged wormholes the total stress--energy tensor can be written as
\begin{equation}
    T_{ab}=T_{ab}^{(m)}+T_{ab}^{(EM)}.
\end{equation}
where $T_{ab}^{(m)}$ and $T_{ab}^{(EM)}$ are the matter and electromagnetic contributions, respectively. The Maxwell stress--energy tensor is
\begin{equation}
    T_{ab}^{(EM)}
    =
    \frac{1}{4\pi}
    \left(
    F_{ac}F_b{}^{c}
    -
    \frac{1}{4}g_{ab}F_{cd}F^{cd}
    \right).
\end{equation}
Contracting with a null vector $k^a$ gives
\begin{align}
    T_{ab}^{(EM)}k^a k^b
    &=
    \frac{1}{4\pi}
    \left[
    F_{ac}F_b{}^{c}k^a k^b
    -
    \frac{1}{4}
    (g_{ab}k^a k^b)
    F_{cd}F^{cd}
    \right] \nonumber\\
    &=
    \frac{1}{4\pi}
    F_{ac}F_b{}^{c}k^a k^b,
\end{align}
where the trace term drops out because $g_{ab}k^a k^b=0$. One obtains
\begin{equation}
    T_{ab}^{(EM)}k^a k^b
    =
    \frac{1}{4\pi}
    (F_{ac}k^c)(F^a{}_{d}k^d)
    \geq 0.
\end{equation}
Thus, the electromagnetic contribution satisfies the null energy condition. The traversability condition still requires
\begin{equation}
    T_{ab}^{(m)}k^{a}k^{b}<0.
\end{equation}
Note that any cosmological--constant--like integration constant, such as in asymptotically dS or AdS wormholes, enters as $\propto g_{ab}$ and therefore does not contribute to the contraction $R_{ab}k^a k^b$, since
\begin{equation}
    \Lambda g_{ab}k^ak^b=0
\end{equation}
vanishes identically due to the null character of $k^a$.
Consequently, claims of traversable wormholes in unimodular gravity supported by non--exotic matter must be interpreted within the constraints imposed by local null geometry. We therefore conclude that unimodular gravity admits no genuinely traversable wormhole solutions supported by ordinary matter satisfying the null energy condition.

\section{Discussion and Conclusions}
\label{sec:discussion}

In the literature, claims of traversable wormholes in unimodular gravity without exotic matter are
typically formulated by imposing the null energy condition directly on the matter stress--energy tensor $T_{\mu\nu}$. In this sense, no redefinition of an effective source is assumed \emph{a priori}.
Our result shows that this assumption is insufficient to assess traversability at the geometric level.

The analysis presented in this work demonstrates that the condition for traversability is entirely
controlled by the local behaviour of null geodesic congruences, as encoded in the null Raychaudhuri
equation. Since unimodular gravity shares the same local geometric structure as general relativity,
the null focusing properties of spacetime are unchanged. Consequently, null defocusing at the throat
necessarily implies violation of the null energy condition, independently of how the field equations
are written or interpreted.

It is important to emphasize that this conclusion is not trivial. Although the null Raychaudhuri equation itself is purely geometric and independent of the gravitational field equations, the physical interpretation of null focusing in terms of matter sources requires a dynamical relation between curvature and stress--energy. Since unimodular gravity modifies the trace sector of Einstein's equations, it was not \emph{a priori} guaranteed that the standard connection between traversability, null focusing, and energy-condition violation would remain unchanged. The present analysis shows that, despite these dynamical differences, the contraction relevant for null focusing, $R_{ab}k^a k^b$, retains exactly the same relation to the matter sector as in general relativity.

Within this framework, constructions claiming traversable wormholes without exotic matter should modify usual kinematical and dynamical definitions, which could fall into one of the following categories:
\begin{enumerate}
\item a redefinition of an effective stress--energy tensor in which the violation of the null energy
condition is shifted from the matter sector to geometric contributions or integration constants;
\item configurations that do not satisfy the covariant traversability criterion derived from the null
Raychaudhuri equation, despite being labelled as wormholes in a coordinate-dependent sense.
\end{enumerate}

A third possible scenario that may be contemplated is the existence of configurations that are traversable only
by timelike observers, while remaining non-traversable for null signals.
From a purely kinematical viewpoint, this would correspond to allowing defocusing of timelike congruences
while maintaining null focusing.
However, the Raychaudhuri equation for timelike geodesic congruences,
\begin{equation}
\dot{\theta}
= -\frac{1}{3}\theta^2
- \sigma_{ab}\sigma^{ab}
+ \omega_{ab}\omega^{ab}
- R_{ab}u^a u^b,
\end{equation}
shows that timelike defocusing requires $R_{ab}u^a u^b<0$ under the usual assumptions.
More importantly, any spacetime region that permits massive particles to traverse a throat while forbidding
null propagation would necessarily violate standard causal ordering, as it would allow timelike signals to
connect events that cannot be linked by null curves.

Such configurations would therefore challenge the causal structure of the spacetime, undermining the role
of null geodesics as the boundary of causal influence.
For this reason, we do not regard ``timelike-only traversability'' as physically admissible within a
consistent relativistic framework.
A systematic analysis of this possibility, and of its implications for causality and energy conditions,
is left for future work.

From the geometric standpoint adopted here, unimodular gravity does not provide a mechanism to evade the
Raychaudhuri obstruction to traversable wormholes. Any genuinely traversable wormhole with twist-free
null generators requires null defocusing at the throat, which in turn implies violation of the null
energy condition along the corresponding null directions. This conclusion holds irrespective of whether
the underlying gravitational dynamics are described by general relativity or by its unimodular
counterpart.

\medskip
\noindent
Our results therefore establish a local no--go theorem for traversable wormholes in unimodular gravity
supported by ordinary matter, clarifying the geometric origin of the obstruction and the precise sense
in which previous claims must be interpreted.
\\

\section*{Acknowledgements}

E.P. acknowledges support from the POSTDOC\_DICYT project 042531CM\_Postdoc, Vicerrectoría de Investigación, Innovación y Creación, Universidad de Santiago de Chile (USACH).

\bibliography{mybib}

\end{document}